\def\zabs{$z_{\rm abs}$}
\def\la{$_<\atop{^\sim}$}
\def\ga{$_>\atop{^\sim}$}

\def\cm2{cm$^{-2}$}
\def\loghi{\mbox{$\log N_{\rm HI}$}}
\def\nhi{\mbox{$N_{\rm HI}$}}
\def\hi{H~{\sc i}}

\def\e{et~al.}

\def\c2{C~{\sc ii}}
\def\c4{C~{\sc iv}}
\def\fe2{Fe~{\sc ii}}
\def\fe3{Fe~{\sc iii}}
\def\mg1{Mg~{\sc i}}
\def\mg2{Mg~{\sc ii}}
\def\zn2{Zn~{\sc ii}}
\def\si2{Si~{\sc ii}}
\def\si4{Si~{\sc iv}}
\def\al2{Al~{\sc ii}}
\def\al3{Al~{\sc iii}}
\def\o1{O~{\sc i}}
\def\o6{O~{\sc vi}}
\def\n1{N~{\sc i}}

\def\approxlt{\mathrel{\spose{\lower 3pt\hbox{$\sim$}}
        \raise 2.0pt\hbox{$<$}}}
\def\approxgt{\mathrel{\spose{\lower 3pt\hbox{$\sim$}}
        \raise 2.0pt\hbox{$>$}}}

\def\ion#1#2{{\rm #1}\,{\sc #2}} 
\documentclass[twocolumn]{aa}
\usepackage{graphicx}
\def\ion#1#2{{\rm #1}\,{\sc #2}}
\usepackage{txfonts}
\begin{document}
   \title{The Most Metal-Rich Intervening Quasar Absorber Known\thanks{Based
   on the UVES observations collected during the DDT ESO prog. ID
   No. 274.A-5030 at the VLT/Kueyen telescope, Paranal, Chile } }

   \author{C. P\'eroux$^1$, V. P. Kulkarni$^2$, J. Meiring$^2$, R. Ferlet$^3$, P. Khare$^4$, J. T. Lauroesch$^5$, G. Vladilo$^6$,  \&
          D. G. York$^7$. }

   \offprints{C. P\'eroux.}

   \institute{$^1$ European Southern Observatory, Garching-bei-M\"unchen, Germany. \email{cperoux@eso.org}\\
$^2$ Dept. of Physics and Astronomy, Univ. of South Carolina, Columbia, USA.\\
$^3$ Institut d'Astrophysique de Paris, UMR7095 CNRS,
Universite Pierre \& Marie Curie, France.\\
$^4$ Dept. of Physics, Utkal University, Bhubaneswar, India.\\
$^5$ Dept. of Physics and Astronomy, Northwerstern University, Evanston, USA.\\
$^6$ Osservatorio di Trieste, Trieste, Italy.\\
$^7$ Dept. of Astronomy and Astrophysics, Univ. of Chicago, Chicago, USA.
             }
\authorrunning{C. P\'eroux et al.}
\titlerunning{The Most Metal-Rich Intervening Quasar Absorber Known}

   \date{Received August 17, 2005; accepted January 3, 2006}

   \abstract{The metallicity in portions of high-redshift galaxies has
been successfully measured thanks to the gas observed in absorption in
the spectra of quasars, in the Damped Lyman-$\alpha$ systems
(DLAs). Surprisingly, the global mean metallicity derived from DLAs is
about 1/10$^{\rm th}$ solar at 0$\la$z$\la$4 leading to the so-called
``missing-metals problem''.  In this paper, we present high-resolution
observations of a sub-DLA system at \zabs=$0.716$ with super-solar
metallicity toward SDSS J1323$-$0021. This is the highest metallicity
intervening quasar absorber currently known, and is only the second
super-solar absorber known to date. We provide a detailed study of
this unique object from VLT/UVES spectroscopy. We derive
[Zn/H]=$+$0.61, [Fe/H]=$-$0.51, [Cr/H]=$<-$0.53, [Mn/H] = $-$0.37, and
[Ti/H] = $-$0.61.  Observations and photoionisation models using the
CLOUDY software confirm that the gas in this sub-DLA is predominantly
neutral and that the abundance pattern is probably significantly
different from a Solar pattern. Fe/Zn and Ti/Zn vary among the main
velocity components by factors of $\sim 3$ and $\sim 35$,
respectively, indicating non-uniform dust depletion. Mn/Fe is
super-solar in almost all components, and varies by a factor of $\sim
3$ among the dominant components. It would be interesting to
observe more sub-DLA systems and determine whether they might
contribute significantly toward the cosmic budget of metals.

    }

   \maketitle
%

\section{Introduction}

Damped Lyman-$\alpha$ systems (DLAs) seen in absorption in the spectra
of background quasars are selected over all redshifts independent of
the intrinsic luminosities of the underlying galaxies. They have
hydrogen column densities, \loghi\ $\ga$ 20.3 and are the major
contributors to the neutral gas in the Universe at high redshifts
(Storrie-Lombardi \& Wolfe 2000; P\'eroux \e\ 2003b). But it has been
suggested that at least some of the \hi\ lies in systems with \hi\
column density below that required by the traditional DLA definition,
in the ``sub-Damped Lyman-$\alpha$ Systems (sub-DLAs)'' with $19.0$ $<$
\loghi\ $<$ 20.3. The DLAs and sub-DLAs offer direct probes
of element abundances over $ > 90 \%$ of the age of the Universe. Zn
is a good probe of the total (gas and solid phase) metallicity, in
DLAs because Zn tracks Fe in most Galactic stars with [Fe/H]$> -$3, it
is undepleted on interstellar dust grains, and the lines of the
dominant ionisation species Zn II are often unsaturated (e.g., Pettini
et al. 1999). Abundances of depleted elements such as Cr or Fe
relative to Zn probe the dust content and the relative abundances can
also yield information about the nucleosynthetic processes (e.g.,
Pettini et al. 1997; Kulkarni, Fall, \& Truran, 1997; P\'eroux et al.,
2002; Khare et al. 2004). A study of the cosmological evolution of the
\hi\ column density-weighted mean metallicity in DLAs (e.g., Kulkarni
\& Fall, 2002) shows surprising results. Contrary to most models of
cosmic chemical evolution (e.g., Malaney \& Chaboyer, 1996; Pei, Fall
\& Hauser, 1999), recent observations indicate at most a mild
evolution in DLA global metallicity with redshift for 0$\la$z$\la$ 4
(Prochaska \e\ 2003; Khare et~al. 2004; Kulkarni, et~al., 2005; and
references therein). Even theoretical models such as
Smoothed-particle-hydrodynamics simulations (Nagamine, Springel, \&
Hernquist, 2004) predict that the true DLA metallicities could be 1/3
solar at z=2.5 and higher at lower redshifts.

Even at z=2.5, making a census of the predicted and observed neutral
comoving densities of gas, $\Omega$, and metals, $\Omega_{\rm Z}$, one
finds that most of the baryons are in the Lyman-$\alpha$ forest but
its metal content is extremely low. The measured value of $\Omega_{\rm
HI}$(DLA) is only a small fraction of $\Omega_{\rm baryons}$ and the
DLA global mean metallicity is about 1/10$^{\rm th}$ solar. The
metallicity of Lyman break galaxies is still poorly constrained; but,
in any case, these objects are known to be star-forming galaxies and
may not be representative of the normal galaxy population. In total,
these three components account for no more than $\approx 10-15$\% of
what we expect to have been produced by z=2.5 (Pettini \e\ 2003;
Bouch\'e \e\ 2005). The missing metals problem in low-redshift DLAs is
even more surprising since the high global star formation rate
estimates at z$>$1.5 (e.g. Madau \e\ 1998) imply that higher
metallicities should be expected at low redshift.

It is possible that $\Omega_Z$(DLA) has been
underestimated. Metal-rich DLAs could obscure quasars due to their
possible high dust content (Fall \& Pei 1993). This may be the reason
for the apparent low metallicity in the DLAs observed in optically
selected quasars (e.g., Fall \& Pei 1993; Boiss\'e et
al. 1998). Recently Vladilo and P\'eroux (2005) have shown that the
fraction of high-redshift DLAs missed due to dust obscuration, could
be up to 50\%, which is consistent with the results of surveys of
radio selected quasars (Ellison \e\ 2001). They have estimated that at
z$\sim$2.3, the real mean metallicity of DLAs could be 5 to 6 times
higher than what is observed, which may help alleviate the missing
metals problems. Indeed, systems at lower redshift may have
significantly more dust at any given metallicity simply because the
dust in these objects has had more time to process the elements.

On the other hand, new lines of evidence are pointing toward lower
\nhi\ quasar absorbers like Lyman Limit Systems (LLS) and sub-DLAs 
being more metal-rich (P\'eroux \e\ 2003a; Jenkins \e\ 2005). The
dust bias, if real, is also likely to be less severe for metal-rich
sub-DLAs as compared to the metal-rich DLAs due to the lower gas and
therefore dust content in the former, for a constant dust to gas
ratio. Thus, the obscuration bias will affect the DLAs at a lower
dust-to-gas ratio as compared to the sub-DLAs. This scenario is
consistent with the recent radio surveys (e.g., Vladilo \& P\'eroux
2005), but still needs to be further quantified observationally.
Indeed, Zn measurements exist for only two sub-DLAs at low $z$: the
marginally super-solar sub-DLA toward Q0058$+$019 with $z_{\rm
abs}$=0.61, \loghi=20.08, and [Zn/H]=$+$0.08 (Pettini \e\ 2000), and
the supersolar sub-DLA toward SDSS J1323$-$0021 with $z_{\rm
abs}$=0.72, \loghi=20.21, and [Zn/H]=$+$0.40 (Khare et al. 2004).
Although Khare et al. reported [Zn/H]=$+$0.40 for this latter
absorber, the modest resolution of their MMT data could not resolve
the Mg I+Zn II $\lambda$ 2026 and Cr II+Zn II $\lambda$ 2062
blends.The extent of line saturation on the derived column densities
was also unclear.  With the goal of addressing these issues with
high-resolution data, we obtained VLT/UVES spectra of this quasar,
which are presented here. These new data are essential to confidently
determine the metallicity by minimising the problem of line
saturation.  Section 2 presents the observational set-up and data
reduction process, while section 3 presents the analysis. Section 4
provides a discussion of the results.

\section{Observations and Data Reduction}

Spectra of SDSS J1323$-$0021 (z$_{\rm em}$=1.390; SDSS mag $g=18.49$)
were acquired in service mode as Director's Discretionary Time (DDT)
on 3$^{\rm rd}$ of March and 13$^{\rm th}$ of March 2005 with the
high-resolution UVES spectrograph mounted on Kueyen Unit 2 VLT
(D'Odorico \e\ 2000). Three exposures of length 4100 sec, 3500 sec and
4100 sec were obtained with standard 390$+$562 settings thus providing
a wavelength coverage of $\sim$3300\AA-4400\AA, 4700\AA-5600\AA\ and
5800\AA-6600\AA.

The data were reduced using the most recent version of the UVES
pipeline to accommodate for the new format of the raw fits file
(version: uves/2.1.0 flmidas/1.1.0).  Master bias and flat images were
constructed using calibration frames taken the closest in time to the
science frames. The science frames were extracted with the ``optimal''
option. The spectrum was then corrected to vacuum heliocentric
reference. The resulting spectra were combined weighting each spectrum
with its signal-to-noise. The final spectra have resolution of 4.7 km
s$^{-1}$ at Zn II $\lambda 2026$. The spectra were divided into 100
{\AA} regions, and each region normalised using cubic spline functions
of orders 1 to 5.

\section{Analysis}

\begin{table*}
\begin{center}
\caption{Parameter fit to the $z_{\rm abs} = 0.716$ sub-DLA model. Velocities
and b are in km s$^{-1}$ and N's are in cm$^{-2}$.} 
\label{t:fit}
\begin{tabular}{l r r r r r r r r r c c}
\hline\hline
         &Vel      & b    &\ion{Mg}{i}   &\ion{Mg}{ii}   &\ion{Fe}{ii}   &\ion{Zn}{ii}   &\ion{Cr}{ii}   &\ion{Mn}{ii}   &\ion{Ti}{ii}   &[Fe/Zn]   &[Mn/Fe]\\ 
\hline
N(X)	 &$-$120.8 & 7.1  &...	         &5.63e12        &5.82e12        &...            &...            &3.75e11        &...            &...       & 0.79  \\
$\sigma$ &...      &...   &...		 &1.66e11	 &4.08e11	 &...		 &...		 &1.26e11	 &...		 &...	    &...    \\
N(X)     &$-$96.9  &16.3  &...	         &3.65e12        &3.64e12        &2.32e12        &$<$2.81e12     &2.97e11        &2.09e11	 &$-$2.65   &0.89   \\	
$\sigma$ &...      &...   &...	         &1.21e11	 &4.31e11	 &6.53e11	 &...	         &1.93e11	 &1.27e11	 &...       &...    \\  
N(X)     &$-$80.1  & 6.2  &3.16e10	 &3.55e12	 &3.92e12	 &...		 &$<$1.69e12	 &2.09e11	 &1.86e11	 &...	    &0.71   \\
$\sigma$ &...      &...   &1.15e10	 &1.36e11	 &3.69e11	 &...		 &...	         &1.23e11	 &8.41e10	 &...	    &...    \\  
N(X)     &$-$62.9  & 5.5  &1.20e11	 &8.13e12	 &4.25e12	 &$<$3.10e11	 &$<$4.77e11	 &2.01e11	 &1.07e11	 &$-$1.71   &0.65   \\   
$\sigma$ &...      &...   &1.23e10	 &3.68e11	 &4.22e11	 &...	         &...   	 &1.01e11	 &8.00e10	 &...	    &...    \\  
N(X)     &$-$51.4  & 5.0  &$<$1.33e10    &$>$2.86e12	 &5.74e12	 &$<$2.56e11	 &...		 &1.44e11	 &$<$6.92e10	 &$-$1.49   &0.38   \\
$\sigma$ &...      &...   &...	         &...	         &1.30e12	 &...	         &...	         &1.19e11	 &...   	 &...	    &...    \\ 
N(X)     &$-$43.0  & 7.4  &4.80e11	 &$>$6.01e13	 &2.41e13	 &1.07e12	 &...		 &1.98e11	 &1.32e11	 &$-$1.49   &$-$0.11\\
$\sigma$ &...      &...   &2.29e10	 &...	         &3.54e12	 &5.18e11	 &...		 &1.43e11	 &9.85e10	 &...	    &...    \\  
N(X)     &$-$13.0  &12.5  &3.04e12	 &$>$2.38e14	 &1.97e14	 &2.57e12	 &...		 &2.40e12	 &4.79e11	 & $-$0.96  &0.07   \\  
$\sigma$ &...      &...   &1.48e11	 &...            &2.54e13	 &6.22e11	 &...		 &3.14e11	 &2.16e11	 &...	    &...    \\ 
N(X)     &3.9      & 30.5 &2.36e12	 &$>$1.48e14	 &4.00e14	 &...		 &...		 &5.33e12	 &$<$4.17e11	 &...	    &0.10   \\ 
$\sigma$ &...      &...   &2.03e11	 &...		 &2.63e13	 &...		 &...		 &5.28e11	 &...            &...	    &...    \\ 
N(X)     &12.0	   & 5.0  &1.44e11	 &...		 &...		 &$<$3.66e11	 &$<$3.20e11	 &...		 &$<$3.89e10	 &...	    &0.01   \\ 
$\sigma$ &...      &...	  &4.81e10	 &...		 &...		 &...  	         &...            &...		 &...		 &...  	    &...    \\
N(X)     & 28.3    &12.0  &2.06e12	 &$>$7.14e13	 &1.15e14	 &1.53e12	 &$<$2.37e12	 &1.28e12	 &3.80e11	 &$-$0.97   &0.03   \\
$\sigma$ &...      &...	  &1.25e11	 &...            &2.94e13	 &7.60e11	 &...            &2.86e11        &1.53e11	 &...  	    &...    \\
N(X)     &44.5     &11.0  &4.76e12	 &$>$2.30e13	 &2.89e14	 &7.76e12	 &...		 &4.34e12	 &4.37e11	 &$-$1.27   &0.16   \\    
$\sigma$ &...      &...	  &4.37e11	 &...		 &5.42e13	 &1.48e12	 &...		 &2.90e11	 &1.22e11	 &...  	    &...    \\
N(X)     &56.6     &6.9	  &8.70e11	 &$>$4.62e13	 &2.83e13	 &$<$2.73e11	 &...		 &6.30e11	 &2.69e11	 &$-$0.83   &0.33   \\    
$\sigma$ &...      &...	  &1.15e11	 &...            &1.82e13        &...            &...	         &1.97e11	 &9.48e10	 &...  	    &...    \\
N(X)     &71.8     &11.9  &1.93e12	 &$>$5.09e14	 &3.12e14	 &2.49e12	 &6.07e12	 &3.62e12	 &1.58e11	 &$-$0.75   &0.04   \\
$\sigma$ &...      &...	  &5.19e10	 &...            &4.29e13	 &7.30e11	 &3.24e12	 &2.22e11	 &1.05e11	 &...  	    &...    \\ 
N(X)     &92.7     &6.4	  &2.42e12	 &$>$2.99e14	 &3.76e13	 &7.35e12	 &$<$1.54e12	 &1.20e12	 &2.09e11	 &$-$2.14   &0.48   \\
$\sigma$ &...      &...	  &8.51e10	 &...		 &4.70e12	 &1.49e12	 &...    	 &1.40e11	 &7.87e10	 &...  	    &...    \\ 
N(X)     &119.3    &8.0	  &7.45e10	 &4.49e12	 &2.33e12	 &5.13e11	 &6.01e12	 &1.98e11	 &1.02e08	 &$-$2.19   &0.91   \\
$\sigma$ &...      &...	  &1.32e10	 &4.90e07	 &2.85e11	 &4.48e11	 &3.11e12	 &1.27e11	 &1.42e07	 &...  	    &...    \\
N(X)     &194.1    &4.6	  &2.44e10	 &1.91e12	 &...		 &...		 &...		 &...		 &...		 &...	    &...    \\  
$\sigma$ &...      &...	  &1.00e10	 &7.56e10	 &...		 &...		 &...		 &...		 &...		  &...	    &...    \\    
\hline 				       			 	  
\end{tabular}			       			 	  
\end{center}			       			 	  
\end{table*}

\begin{figure*} 
\centering 
\includegraphics[width=.65\textwidth]{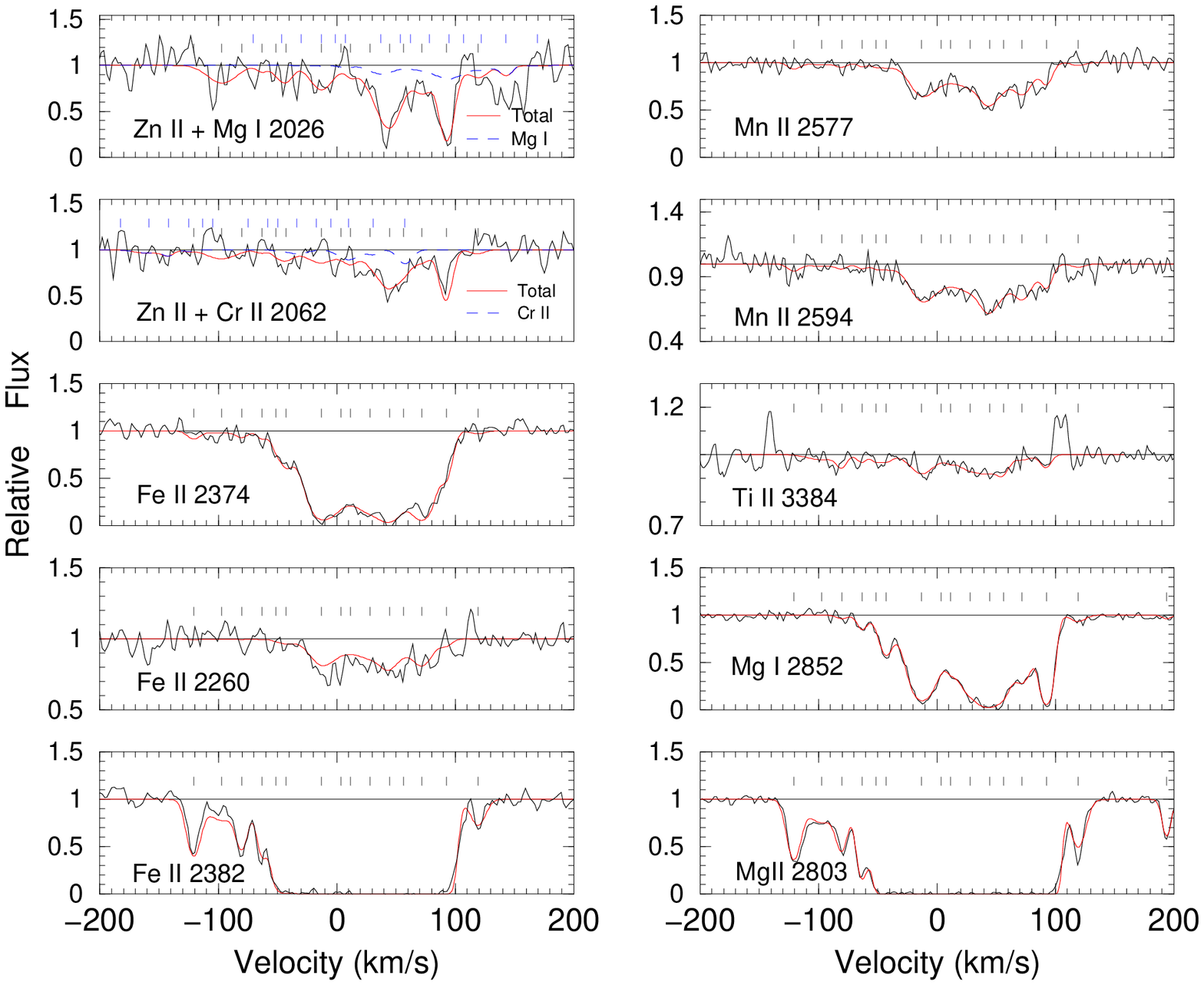}
\caption{Velocity plots of some of the element lines detected at $z_{\rm
abs}=0.716$ toward SDSS J1323$-$0021. The black tick-marks indicate
positions of the components and the red curves denote the best-fit
profiles. In the two top left panels, the upper set of blue tickmarks
indicate positions of the components of Mg I $\lambda 2026$ and Cr II
$\lambda 2062$, respectively, whilst the blue curves show their
contributions to the absorption profiles. Note that different
ordinates are used on the sub-frames. }
\label{f:fit}
\end{figure*}

Several lines of Zn II, Cr II, Fe II, Mn II, and Ti II were detected
at z$_{\rm abs}$=0.716.  Fe I, Zn I, and Co II were not detected. The
column densities were estimated by fitting multi-component Voigt
profiles to the observed absorption lines using the program FITS6P
(Welty, Hobbs, \& York 1991) that evolved from the code used by
Vidal-Madjar et al. (1977). FITS6P minimizes the $\chi^{2}$ between
the data and the theoretical Voigt profiles convolved with the
instrumental profile. The atomic data were adopted from Morton
(2003).

The absorption profiles show a complex velocity structure with a total
of 16 components needed. The velocity and Doppler b parameters of the
various components were estimated from the Mg I, Mg II, and Fe II
lines. The component at 12 km s$^{-1}$ is negligible in most species
except Mg I, Zn II and Cr II. The component at 194 km s$^{-1}$ is
negligible in all species but Mg II. The component at $-$121 km
s$^{-1}$ is detected in Mg II, Fe II, and Mn II, but not in Mg I, Zn
II, Cr II, or Ti II.  The best-fit column densities in the individual
components and their uncertainties were estimated assuming the same
fixed b and v values for all species in a given component
(Figure~\ref{f:fit}). The results of the profile fitting analysis are
summarised in Table~\ref{t:fit}. Column densities in the few weak
components that could not be well-constrained due to noise are marked
with ``...''; their contributions to the total column densities
(listed in Table~\ref{t:ab}) are negligible. As an additional check,
we also estimated the total column densities using the apparent
optical depth (AOD; Savage \& Sembach 1991) method for the various
detected lines and obtained results consistent with those from the
profile-fitting method. The agreement was within 0.01 dex for Mg I and
Mn II, within 0.1 dex for Fe II and Zn II, and within 0.15 dex for Ti
II.

The Mg I $\lambda 2026.5$ contribution to the Zn II $\lambda 2026.1$
line was estimated using the component parameters for Mg I derived
from the Mg I $\lambda 2852$ profile. This contribution, indicated by
a dashed blue curve in the top left panel of Fig~\ref{f:fit}, was a
small fraction of the observed strength of the $\lambda 2026$
line. The remaining part of the $\lambda 2026$ line was fitted with a
15-component model for Zn II, using the same $b$ values and
velocities, but varying the column densities in the individual
components. The Zn II fit thus obtained was used to estimate the Zn II
contribution to the $\lambda 2062$ line. The remaining part of the
$\lambda 2062$ line was fitted with a 15-component model for Cr II,
using the same set of $b$ values and velocities.  This was the only
available estimate for Cr II, since the Cr II $\lambda \lambda$ 2056,
2066 lines lie in noisy regions and are undetected.  The relatively
large errors in the Cr II column densities arise from the noisy nature
of the $\lambda 2062$ line.  The Fe II column densities in the
components at low velocities were constrained by using the Fe II
$\lambda \lambda 2260, 2374$ lines, since the stronger Fe II lines are
saturated. The weaker Fe II components at high positive and negative
velocities were constrained in column density using the $\lambda
\lambda 2374, 2382, 2600$ lines, since these components are poorly
constrained by the weaker $\lambda \lambda 2260,2374$ lines.  The Mg
II $\lambda 2796, 2803$ profiles were fitted together, but provide
only a lower limit to \mg2\ owing to saturation in the central
components. The relative abundances were calculated using solar
abundances from Asplund et al. (2005), adopting the mean of
photospheric and meteoritic values for Mg, Ti, Cr, Fe, Zn, and the
meteoritic value for Mn.

One concern is that sub-DLAs may be partially ionised in H,
artificially enhancing the ratio of Zn II to H I, for instance. To
investigate the ionisation corrections, we used the CLOUDY software
package (version 94.00, Ferland 1997) and computed photoionisation
models assuming ionisation equilibrium and a solar abundance
pattern. We thus obtained the theoretical column density predictions
for any ionisation state of all observed ions as a function of the
ionisation parameter U. Our findings confirm the observations: from a
comparison of the observed and theoretical Mg II/Mg I ratios, we
deduced that the gas in this sub-DLA is predominantly neutral ($\log U
<- 5$) and the overall abundance pattern is probably not solar.  This
latter point is also clear from the relative abundances listed for
each component in Table~\ref{t:fit}. It should be pointed out
that there are no third ionisation stage detected in the system under
study. Nevertheless, we do have Ti II, which has the same ionisation
potential as H I. The fact that Ti is not suppressed compared to Fe or
Mn also implies that the gas is neutral with low ionisation
parameter.

\begin{table}
\begin{center}
\caption{Summary of total abundances.} 
\label{t:ab}
\begin{tabular}{l c l c }
\hline\hline
Id&$\log N_{\rm total}$&A(X/N)$_{\sun}$&[X/H]$^*$\\
\hline
\ion{Mg}{i}    &$13.26\pm0.01$      &...   &...  \\
\ion{Mg}{ii}   &$>15.15$ &$-$4.47 &$>-$0.58\\
\ion{Fe}{ii}   &$15.15 \pm 0.03$	 &$-$4.55 &$-$0.51 $\pm 0.20$\\
\ion{Zn}{ii}   &$13.43 \pm 0.05$	 &$-$7.40 &$+$0.61$\pm 0.20$\\
\ion{Cr}{ii}   &$<13.33$	 &$-$6.37 &$<-$0.52\\
\ion{Mn}{ii}   &$13.31 \pm 0.02$&$-$6.53 &$-$0.37$\pm 0.20$\\
\ion{Ti}{ii}   &$12.49 \pm 0.11$&$-$7.11 &$-$0.61$\pm 0.22$\\
\hline 
\end{tabular}
\end{center}
$^*$ The error bars on [X/H] include the errors in log $N(X)$ and
\loghi.
\end{table}

\section{Discussion and Conclusions}

Table~\ref{t:ab} lists the abundances, using
$\loghi=20.21^{+0.21}_{-0.18}$ (Khare \e\ 2004) that we derived from
Voigt profile fitting of the damped Ly-$\alpha$ line in the
publicly available HST/STIS spectrum of SDSS J1323$-$0021 (program GO
9382; PI: Rao). Rao et al. (2005) obtained
$\loghi=20.54^{+0.16}_{-0.15}$ from the same data set. We use the
former value since it gives a smaller residual with respect to the
data and therefore regard this absorber as a sub-DLA. For either \nhi,
the strength of the Zn II lines detected in our UVES spectrum implies
a super-solar metallicity. Using the standard definition: $[X/H] =
\log [N(X)/N(H)]_{DLA}- \log [N(X)/N(H)]_{\odot}$, we find
[Zn/H]=$+$0.61.

In principle, if Mg I $\lambda$ 2852 were substantially saturated, the
contribution of Mg I $\lambda$ 2026 could be higher than our best-fit
estimate. However, based on our apparent optical depth measurements
and profile-fitting results, we estimate that it would take $\sim$8
times more total Mg~{\sc i} than the value derived from the $\lambda
2852$ line to contribute the entirely of the $\lambda 2026$ line. Such
a high value of Mg~{\sc i} can be ruled out by the observed profile of
the $\lambda 2852$ line. (Of course, such a scenario would also be
inconsistent with the observed strength and profile of the $\lambda
2062$ line.) To understand this issue in more detail, we estimated the
maximum Mg~{\sc i} in the dominant components that would still give
the shape of the Mg I $\lambda$2852 profile consistent with the
observed profile within the noise level in the continuum. This maximum
total Mg~{\sc i} $= 2.7e13$ is about $50 \%$ larger than the best-fit
value listed in Table 2. Putting this maximum Mg I model in the
$\lambda 2026$ line, the corresponding total Zn~{\sc ii} needed to fit
the remaining part of the $\lambda$ 2026 line would be 2.5e13, lower
by $< 10 \%$ from our best-fit value. Thus [Zn/H] is at least
$>+$0.59, indicating that our result would not be affected much by
saturation of Mg I $\lambda$ 2852. Finally, as an additional check, we
also estimated the maximum contribution of Cr II 2062 by rebinning our
spectrum by factors of 10 or 20, measuring the upper limit for Cr
II$\lambda$ 2056 in the rebinned spectrum. We then spread this upper
limit for N(Cr II) over the 2062 profile, using the velocity model
derived from the combination of the lines. We assume the same Fe/Cr
ratio in all components, taking the percentage of Cr II in each
components relative to total N(CrII) summed over all components to be
the same as the corresponding percentage of Fe II in that component.
Fitting the remaining part of the $\lambda 2062$ line with a
15-component model of Zn II, we estimated $N_{\rm Zn II} > 2.05e13$,
i.e. $[Zn/H] > 0.50$.

The abundances of Fe, Cr, Mn, and Ti lie in the range of $-$0.4 to
$-$0.6 dex, and indicate that this absorber is not only metal-rich,
but also very dusty. Using the model from Vladilo (2004), we find that
95\% of the Fe is in dust phase and the total metallicity is even
slightly higher than 0.6 dex. This sightline also shows substantial
reddening compared to the SDSS quasar composite ($\Delta (g-i) =
0.47$; Khare et al. 2004). Considering only the better-determined
components between $-$13 and 100 km s$^{-1}$, Fe/Zn varies by a factor
of $\sim3$ and Ti/Zn varies by $\sim$35. [Mn/Fe] varies by $\sim 3$,
but indicates a super-solar Mn abundance with respect to Fe in all
components. The relative abundance of Mn with respect to Fe is not
expected to exceed the solar value as Mn, unlike Fe, has an odd atomic
number. However, [Mn/Fe]$>$0 is often seen in the Galactic
interstellar gas due to the stronger depletion of Fe on to dust
grains.

To summarise, our high resolution VLT/UVES data have allowed us
to alleviate the saturation issue in Zn II and Cr II lines and
therefore unambiguously prove the super-solar metallicity of the
sub-DLA at $z_{\rm abs}$=0.716 toward SDSS J1323$-$0021. If the dust
obscuration bias for DLAs is indeed significant, as proposed by
Vladilo and P\'eroux (2005), sub-DLA systems such as the one reported
here could be better probes of dusty regions with significant past
star formation (e.g. Lauroesch \e\ 1996; York \e\ 2006), as similar
DLA systems will be missed due to dust obscuration.  On the other
hand, it is also, possible to envisage a scenario where the dust
obscuration is not very significant in quasar absorbers (as is
indicated by the rising spectrum of gamma-ray burst afterglows). In
this scenario it is possible that the sub-DLA systems may indeed be
more metal-rich as compared to DLAs as indicated by observations of
P\'eroux \e\ (2003a) and by the observations of super-solar
metallicity in such systems presented here and in Pettini et al
(2003). With the large-scale spectroscopic surveys of quasars
currently underway (e.g. the SDSS, York \e\ 2000, York \e\ 2001), such
metal-rich sub-DLA systems may be found in large numbers. If future
observations indeed find such systems, sub-DLAs may contribute
significantly to the overall global metallicity.

\begin{acknowledgements}
We are grateful to ESO director, Catherine Cesarsky, for time
allocation to this DDT program, and to the VLT staff for carrying out
our observations in service mode.  VPK and JM acknowledge support from
the U. S. National Science Foundation grant AST-0206197.
\end{acknowledgements}


\begin{thebibliography}{}

\bibitem[]{} Asplund, M., Grevesse, N. \& Sauval, A. J., ASP Conference Series, 2005. Ed Bash \& Barnes, 336, 25.

\bibitem[]{} Bouch\'e, N., Lehnert, M. D. \& P\'eroux, C., 2005, MNRAS, 364, 319. 

\bibitem[]{} Boiss\'e, P., Le Brun, V., Bergeron, J. \& Deharveng, J-M., 1998,  1998, A\&A, 333, 841.


\bibitem[]{} D'Odorico, S., Cristiani, S., Dekker, H., Hill, V., Kaufer, A., Kim, T. \& Primas, F., 2000, SPIE, 4005, 121

\bibitem[]{} Ellison, S. L., Yan, L., Hook, I. M., Pettini, M., Wall, J. V. \& Shaver, P., 2001, A\&A, 379, 393.

\bibitem[]{} Fall, M. \& Pei, Y., 1993, ApJ, 402, 479.

\bibitem[]{} Ferland, G.J., 1997, Hazy, a brief introduction to CLOUDY

\bibitem[]{} Jenkins, E. B., Bowen, D. V., Tripp, T. M. \&  Sembach, K. R., 2005, ApJ, 623, 767.

\bibitem[]{} Khare, P., Kulkarni, V. P., Lauroesch, J. T., York, D. G., 
Crotts, A. P. S. \& Nakamura, O., 2004, ApJ, 616, 86.

\bibitem[]{} Kulkarni, V. P., Fall, S. M. \& Truran, J. W., 1997, ApJ, 484,
L7.

\bibitem[]{} Kulkarni, V. P. \& Fall, S. M., 2002, ApJ, 580, 732.

\bibitem[]{} Kulkarni, V. P., Fall, S. M., Lauroesch, J. T., York, D. G., 
Welty, D. E., Khare, P. \& Truran, J. W., 2005, ApJ, 618, 68.

\bibitem[]{} Lauroesch, J. T., Truran, J. W., Welty, D. E. \& York, D. G., 
1996, PASP, 108, 641.

\bibitem[]{} Madau, P., Pozzetti, L. \& Dickinson, M., 1998, ApJ, 498, 106.

\bibitem[]{} Malaney. R. A. \& Chaboyer, B., 1996, ApJ, 462, 57.

\bibitem{M03} Morton, D. C. 2003, \apjs, 149, 205.

\bibitem[]{} Nagamine, K., Springel, V. \& Hernquist, L., 2004, MNRAS, 348, 
435.

\bibitem[]{} Pei, Y. C., Fall, S. M. \& Hauser, M. G., 1999, ApJ, 522, 604.

\bibitem[]{} P\'eroux, C., McMahon, R., Storrie-Lombardi, L., \&
Irwin, M., 2003a, MNRAS, 346, 1103.

\bibitem[]{} P\'eroux, C., Dessauges-Zavadsky, M., D'Odorico, S., Kim, T. S. 
\& McMahon, R., 2003b, MNRAS, 345, 480.

\bibitem[]{} P\'eroux, C., Petitjean, P., Aracil, B. \& Srianand,
A., 2002, New A., 7, 577.

\bibitem[]{} Pettini, M., King, D. L., Smith, L. J. \& Hunstead, R. W., 
1997, ApJ, 478, 536.

\bibitem[]{} Pettini, M., Ellison, S. L., Steidel, C. C. \& Bowen, D. V., 
1999, ApJ, 510, 576.

\bibitem[]{} Pettini, M., 2003, XIII Canary Islands Winter School, ed Esteban,
L\'opez, Herrero, S\'anchez.

\bibitem[]{} Pettini, M., Ellison, S. L., Steidel, C. C., Shapley, A. E. 
\& Bowen, D. V., 2000, ApJ, 532, 65.

\bibitem[]{} Prochaska, J., Gawiser, E., Wolfe, A. M., Castro, S. 
\& Djorgovski, S. G., 2003, ApJ, 595L, 9.

\bibitem{} Rao, S. M., Turnshek, D. A., \& Nestor D. B. 2005, ApJ, 129, 9.

\bibitem{R05} Rao, S. M., Turnshek, D. A., \& Nestor D. B. 2005 
{\it in press}.

\bibitem[]{} Savage, B. D. \& Sembach, K. R., 1991, ApJ, 379, 245.

\bibitem[]{} Storrie-Lombardi, L. \& Wolfe, A., 2000, ApJ, 543, 552.

\bibitem{VM77} Vidal-Madjar, A.,  Laurent, C., Bonnet, R. M., 
\& York, D. G. 1977, ApJ, 211, 91.

\bibitem[]{} Vladilo, G., 2004, A\&A, 421, 479.

\bibitem[]{} Vladilo, G. \& P\'eroux, C., 2005, A\&A, 444, 461.

\bibitem{WHY91} Welty, D. E., Hobbs, L. M., \& York, D. G. 1991, ApJS, 
75, 425.

\bibitem[]{} York, D. G. and the SDSS Collaboration, 2000, AJ, 120, 1579.

\bibitem[]{} York, D. G. and the SDSS Collaboration, 2001, BAAS, 198, 78.05.

\bibitem[]{} York, D. G. et al., 2006, {\it in preparation}.

\end{thebibliography}
\end{document}